\documentclass[12pt]{article}
\usepackage{amssymb,amsmath,cite}
\newcommand{\be}{\begin{equation}}
\newcommand{\ee}{\end{equation}}
\newcommand{\bea}{\begin{eqnarray}}
\newcommand{\eea}{\end{eqnarray}}

\long\def\@makefntext#1{\parindent 0cm\noindent
\hbox to 1em{\hss$^{\@thefnmark}$}#1}
\def\Z#1{_{\lower2pt\hbox{$\scriptstyle#1$}}}

\begin{document}
\begin{titlepage}
\vspace{.5in}
\begin{flushright}
22 July 2009\\
\end{flushright}
\vspace{.5in}
\begin{center}
{\large\bf Accelerating universe from warped extra dimensions }\\
\vspace{.4in}
{I{\sc shwaree} ~P.~N{\sc eupane}\footnote{\it email: ishwaree.neupane@canterbury.ac.nz}\\
       {\small\it Department of Physics \& Astronomy}\\
       {\small\it University of Canterbury, Private Bag 4800}\\
       {\small\it Christchurch 8041, New Zealand}\\}
\end{center}

\vspace{.5in}
\begin{center}
{\large\bf Abstract}
\end{center}
\begin{center}
\begin{minipage}{4.7in}

{\small Accelerating universe or the existence of a small and
positive cosmological constant is probably the most pressing
obstacle as well as opportunity to significantly improving the
models of four-dimensional cosmology from fundamental theories of
gravity, including string theory. In seeking to resolve this
problem, one naturally wonders if the real world can somehow be
interpreted as an inflating de Sitter brane embedded in a
higher-dimensional spacetime described by warped geometry. In this
scenario, the four-dimensional cosmological constant may be
uniquely determined in terms of two length scales: one is a scale
associated with the size of extra dimensions and the other is a
scale associated with the expansion rate of our universe. In some
specific cases, these two scales are complementary to each other.
This result is demonstrated here by presenting some explicit and
completely non-singular de Sitter space dS$_4$ solutions of vacuum
Einstein equations in five and ten dimensions.}

\medskip


\end{minipage}
\end{center}
\end{titlepage}
\addtocounter{footnote}{-1}

Experimental evidence for a non-vanishing cosmological energy
density in today's universe is among the most significant
discoveries in modern cosmology~\cite{supernovae}. This result
together with an inflationary epoch required to solve the horizon
and flatness problems of the big bang cosmology needs to be
understood in the framework of fundamental theories of gravity.
One wonders whether one should look for an alternative to a true
cosmological constant.

God might have played tricks with physicists -- as Einstein
believed. One of those tricks appears to be with the observed
value of the vacuum energy density $\Lambda_4\sim 10^{-47}~({\text
GeV})^4$, which is extremely small by particle physics
standard~\cite{Weinberg}. Faced with this enigma, one way out is
an anthropic principle. This entails abandoning the quest for a
conventional scientific explanation to cosmological constant and
interpreting the smallness of the cosmological vacuum energy
density as a suitable variable of our local needs.

As the problem of cosmological constant could involve quantum
gravity, higher dimensional physics is the only framework for
addressing it, at least with our present state of knowledge.
Assuming that the physics in higher spacetime dimensions defines
the framework of low-energy effective four-dimensional field
theory and a unique answer to the vacuum energy density, there
will be a unique prediction for the cosmological constant in
today's universe. I argue, in this paper, that the vacuum energy
density of our spacetime may be determined by just two parameters:
one is a scale related to the size of extra dimensions and the
other is a scale related to the present size and/or the Hubble
expansion rate of the universe. This paper, building on some new
ideas, provides a key conceptual consolidation.

The basic idea can be illustrated with a five-dimensional metric
of the general form
\begin{equation}\label{sol-5Da}
ds_5^2 = e^{- 2\mu |z|} \left(\hat{g}_{\alpha\beta} dX^\alpha
dX^\beta + e^{-2\lambda}\, {dz}^2\right).
\end{equation}
Here, $X^\alpha$ are the usual spacetime coordinates ($\alpha,
\beta =0, 1, 2, 3$), $e^{-\lambda}$ is a length scale associated
with the size of extra space and $z\ge 0$; there is some sort of
`brane' at the $z=0$ boundary of spacetime. If
$\hat{g}_{\alpha\beta}=\eta_{\alpha\beta}$, in which case the
`brane' is flat or four-dimensional Minkowski space, this metric
describes a portion of five-dimensional anti de Sitter space, with
negative cosmological constant, as in Randall-Sundrum braneworld
models~\cite{RS1,RS2}.

The problem of the vacuum energy density or cosmological constant
-- why it is extremely small by particle physics standard --
really only arises in a dynamical spacetime. It is perhaps natural
to assume that the five-dimensional cosmological constant
vanishes. One may ask, can a four-dimensional graviton be
`localized' near $z=0$ in a flat 5D Minkowski spacetime?
Localization of gravity means a finiteness of effective
four-dimensional Newton's constant. I shall argue, in this paper,
this is possible provided that we live in a universe that is not
static rather inflating.

The metric (\ref{sol-5Da}) becomes an exact solution of 5D vacuum
Einstein equations when
\begin{equation}\label{FRW}
\hat{g}_{\alpha\beta} dX^\alpha dX^\beta =   -dt^2+
a^2(t)\left[\frac{ dr^2}{1-k (r/r_0)^2}+ r^2
(d\theta^2+\sin^2\theta d\phi^2)\right],
\end{equation}
and the scale factor of the universe $a(t)$ is given by
\begin{equation}\label{main-sol-scale1}
a(t)= \frac{1}{2} \exp\left(\mu t \, e^\lambda \right)
+\frac{\tilde{k}\, e^{-2\lambda}}{2 \mu^2}\, \exp\left( - \mu t\,
e^\lambda\right),
\end{equation}
where $\mu$ is an integration constant, $\tilde{k}\equiv k/r_0^2$
and $k=0, \pm 1$. Here, both the scale factor $a(t)$ and the warp
factor $W(z)\equiv e^{-\mu |z|}$ are defined up to the rescaling
$t\to t+t_0$ and $z\to z+z_0$. Clearly, with $\lambda\ne 0$, the
scale factor and the warp factor will have different slopes. The
universe naturally inflates when $\mu\ne 0$ or when the warp
factor $e^{-\mu z}$ is not constant. In an open universe ($k=-1$),
there is a big-bang type singularity at $\mu t=-(\lambda+ \ln
\left(\mu r_0 \right)) e^{-\lambda}$, whereas in a closed ($k=1$)
or spatially flat ($k=0$) universe, the scale factor is always
non-zero. This result is desirable because the generic singularity
of a time-dependent solution of general relativity is generally
unacceptable -- as it may not have any quantum interpretation.

A choice like $\lambda=0$ could simply lead to an erroneous
observation that in the present universe, since
$H_0=\dot{a}/a=\mu\sim 10^{-60} M_{Pl}$ (with $k=0$), the warping
of an extra space may not be sufficiently strong as to make the
extra space geometrically compact. In fact, the coefficient
$e^{\lambda}$ multiplying the extra-dimensional metric should be
fixed using some phenomenological constraints, such as, the
Kaluza-Klein mass scale associated with the size of extra
dimensions becomes not too small, such as $m_{KK} \sim e^\lambda
\gtrsim {\rm TeV}$ (in natural units).

One proceeds by specifying boundary conditions such that the warp
factor is regular at $z=0$, where one places a 3-brane with brane
tension $T_3$, and well behaved at infinite distances from the
brane. The classical action describing this set-up is
\begin{equation}
S= \frac{M\Z{5}^3}{2} \int_B d^5{x} \sqrt{-g\Z{5}}\,R_{(5)} +
\frac{M\Z{5}^3}{2} \int_{\partial B} d^4 x\sqrt{-g\Z{b}}(- T_3),
\label{main-action2}
\end{equation}
where $M_5$ is the fundamental 5D Planck scale and $g_{b}$ is the
determinant of the metric $g_{ab}$ evaluated at the brane's
position. Einstein's equations are also solved at the brane's
position when
\begin{equation}\label{sol-with-T3}
T_3= 12\mu \,e^{\lambda}.
\end{equation}
The brane tension is induced not by a bulk cosmological constant
but by the curvature related to the expansion of the physical
$3+1$ spacetime, which vanishes only in the limit $\mu \to 0$. In
the same limit, the scale factor $a(t)\to {\rm const}$ (when
$k=0$) and the warp factor $W(z)=e^{-\mu z}\to 1$. This defines a
flat Minkowski vacuum. With a nonzero brane tension the expansion
of the universe must accelerate eventually. In a wide class of
models, this suggestion has some startling implications that I
would like to emphasize.

First, the physical universe can naturally inflate due to the
warping of extra space as it generates in the four-dimensional
effective theory a cosmological constant-like term $\Lambda_4$.
This follows by substituting equation~(\ref{sol-5Da}) into the
classical action~(\ref{main-action2}). We shall focus here on the
5D curvature term from which one can derive the scale of
gravitational interactions:
\begin{equation}\label{effective-S}
S_{\rm eff} \supset \frac{M_5^3}{2} \int d^4 x
\sqrt{-\hat{g}\Z{4}} \int e^{-\lambda} \, dz\, e^{-3\mu |z|}
\left(\hat{R}_4 - 2\Lambda_4 \right),
\end{equation}
where the four-dimensional cosmological constant is given by
\begin{equation}\label{CC-sol}
\Lambda_4 = {6 \mu^2}\,e^{2\lambda}.
\end{equation}
Second, as advertised above, the vacuum energy density is uniquely
determined in terms of two length scales: one is a scale
associated with the size of extra space, i.e. $e^\lambda$ and the
other is a scale associated with the expansion rate of the
universe that we live in or the slope of the warp factor, i.e.
$\mu$.

A strict cosmological constant could be quite unrealistic as it
introduces effective length and time cut-offs. For the scale given
by~(\ref{main-sol-scale1}), it is determined by a product of two
scales, one associated with the slope of the warp factor (or the
expansion rate of the physical universe) and the other associated
with the size of extra dimension (see for a related
discussion~\cite{Wang-etal}), so there is no any such pathology at
least in spatially flat and closed universes. The model also
exhibits certain features 5D braneworld model discussed earlier
in~\cite{DGP} where the cosmic acceleration may be sourced by the
scalar curvature term on the brane (or 4d hypersurface).

Third, from equation~(\ref{effective-S}),
 we find that the relation between four- and five-dimensional
 effective Planck masses is given by
\begin{equation}
 M_{4}^2 = {M_5^3\, e^{-\lambda} }\int_{-z_1}^{z_1} d{z}\,e^{-3\mu |z|}=
 \frac{2 M_5^3 e^{-\lambda} }{3\mu} \left[1-e^{-3\mu z\Z{1} }\right].
\end{equation}
There is a well-defined value for $M_{4}$, even in the $z_1\to
\infty$ limit. In the limit $\mu\to 0$, a new dimension of
spacetime opens up, since $M_4^2 \to \infty$ (or $G_N\to 0$) as
$z\Z{1}\to \infty$. All the above results are quite generic and
hold also when the spacetime dimensions $D>5$. Based on earlier
braneworld no-go theorems~\cite{Gibbons-84,Malda-Nunez}, it seemed
almost impossible for these results to be true, but with what we
presently understand~\cite{Ish:09a,Ish:09c}, inflationary
cosmology is possible for a wide class of metrics without
violating even the weakest form of a positive energy condition,
which is that the stress tensor $T_{AB}$ obeys $\xi^A \xi^B
T_{AB}\ge 0$ for any light-like vector ${\xi^A}$ in $D=5$.

Superstring theory predicts that space has six more spatial
dimensions, other than the usual three large dimensions. So from
the viewpoint of string theory, we find it more natural to
generalize the above discussion, demanding the existence of
additional five spatial dimensions, which can be topologically
compact. This provides us the relevant framework of a
ten-dimensional supergravity model characterized by the following
warped metric,
\begin{equation}\label{10d-gen-anz}
ds\Z{10}^2 = e^{2A(z)} \hat{g}_{\alpha\beta} dX^\alpha dX^\beta +
e^{\alpha\Z{0} A(z)}\,e^{-2\lambda} \left(f_1(z) \,dz^2
+\alpha\Z{1}\, f_2(z)\,ds^2\Z{X_5}\right),
\end{equation}
where $\alpha\Z{0}$ and $\alpha\Z{1}$ are some constants. The
functions $f_1(z)$ and $f_2(z)$ may be chosen to be
\begin{equation}\label{f1-and-f2}
f_1(z)=\sinh^2(z+z_0), \quad f_2(z)= \cosh^2 (z+z_0).
\end{equation}
The five-dimensional base space $X_5$ can be a usual five-sphere
$S^5$ or some other compact (Calabi-Yau) manifolds. A particularly
interesting example of $X_5$ is the Einstein space $(S^2\times
S^2)\rtimes S^1$:
\begin{equation}
ds^2\Z{X_5} = \frac{1}{9} \left(d\psi+\cos\theta_1 d\phi_1
+\cos\theta_2 d\phi_2\right)^2 + \frac{1}{6} \sum_{i=1}^{2}
\left(d\theta_i^2+ \sin\theta_i^2 d\phi_i^2\right),
\end{equation}
where $(\theta_1, \phi_1)$ and $(\theta_2, \phi_2)$ are
coordinates on each $S^2$ and $\psi$ is the coordinate of a $U(1)$
fiber. With the choice of $f_1$ and $f_2$ as in~(\ref{f1-and-f2})
the radius of the $X_5$ never vanishes. The 6d Ricci scalar
curvature is $R_6=20(1-\alpha\Z{1})/[\alpha\Z{1}\cosh^2(z+z_0)]$,
which is smooth everywhere. One also notes that the 6d part of the
metric becomes Ricci flat only when $\alpha\Z{1}=1$.

Let us first consider the case $\alpha_0< 2$; the $\alpha\Z{0}=2$
case will be treated separately. One chooses the coefficient
$\alpha\Z{1}$ in equation~(\ref{10d-gen-anz}) satisfying
$8\alpha\Z{1}=(2-\alpha\Z{0})^2$, as required to allow a
four-dimensional de Sitter solution in all three cases, i.e.
$R_6>0$ or $R_6=0$ or $R_6<0$. It is not difficult to see that the
10d vacuum Einstein equations are explicitly solved for
\begin{equation}\label{main-sol-nonsin}
A(z)=\frac{1}{2-\alpha\Z{0}}\ln
\left(\frac{3\mu^2(2-\alpha\Z{0})^2 \cosh^2(z+z_0)}{32}\right),
\quad a(t)=\frac{1}{2} \exp\left({\mu t} e^{\lambda}\right)+
\frac{\tilde{k} e^{-2\lambda}}{2\mu^2} \exp\left(-\mu t
e^{\lambda}\right).
\end{equation}
This result is quite remarkable. First, the solution is free from
both space-like and time-like singularities. Second, it provides
an explicit example of a warped compactification on a
four-dimensional de Sitter space dS$_4$ just by solving the 10d
vacuum Einstein equations.

In ~\cite{Gibbons-Hull}, Gibbons and Hull constructed warped
supergravity models that allow solutions without (external brane)
sources. The solutions presented in~\cite{Gibbons-Hull} are,
however, singular at the center ($r=0$) where the radius of $X_5$
vanishes, whereas the solution presented above is regular
everywhere. Moreover, the Randall-Sundrum mechanism, i.e.
localization of gravity due to a non-singular delta function
source, is applicable to the present model.

In order to satisfy the field equations at $z=0$, where one places
a 8-brane with brane tension $T_8$, one writes the 10d Einstein
equations in the form
\begin{equation}\label{10d-boundary}
G_A^B= {T}_8\,\frac{\eta_M^N }{\sqrt{g_{zz}}}\,\delta_A^{M}
\delta^B_N\, \delta(z),
\end{equation}
where $(M, N)=(t, x_i, \theta_i, \phi_i, \psi)$. The 8-brane (with
tension $T_8$) is extended in all of the $x^\mu$ directions and
along the $X_5$ space. From eqs.~(\ref{10d-gen-anz}) and
(\ref{main-sol-nonsin}) one explicitly finds
\begin{eqnarray}\label{10d-brane-sol}
{G_z^z}|\Z{z=0}=0, \quad G_M^N|\Z{z=0}= \frac{32}{(2-\alpha\Z{0})}
\frac{e^{2\lambda}\,e^{-\alpha\Z{0}
A\Z{0}}}{\sinh{z}\Z{0}\cosh{z}\Z{0}}\,\delta(0)\eta_M^N,
\end{eqnarray}
where
\begin{equation}
e^{A\Z{0}}\equiv \left(\frac{3\mu^2 (2-\alpha\Z{0})^2 \cosh^2
z\Z{0}}{32}\right)^{1/(2-\alpha\Z{0})}.
\end{equation}
From eqs.~(\ref{10d-boundary}) and (\ref{10d-brane-sol}), it
follows that
\begin{equation}
T_8=\frac{32}{2-\alpha\Z{0}}\frac{e^\lambda}{\cosh{z}\Z{0}}\,e^{-\alpha\Z{0}
A\Z{0}/2}.
\end{equation}
Furthermore, from (\ref{10d-gen-anz}), one obtains
\begin{equation}
{}^{(10)}R_{\alpha\beta}(x,z)={}^{(4)}R_{\alpha\beta}(x)
-\hat{g}_{\alpha\beta}\,e^{2\lambda}\left(3\mu^2
+\frac{3(2-\alpha\Z{0})\mu^2}{8}\,\coth(z+z_0)\,\delta(z)\right).
\end{equation}
If one integrates this equation on both sides over the internal
manifold, then we clearly find that $^{(4)}R_{00}\equiv
-3\ddot{a}/a<0$ is possible, without violating the 10d strong
energy condition.

The case $\alpha\Z{0}=2$ is special in the regard that in this
case the conformal warp factor $e^{2A(z)}$ multiplying the 4d and
the 6d parts of the metric are the same. The 10d metric reads
\begin{equation}\label{10d-soln1}
ds\Z{10}^2 = e^{2A(z)} \left[ \hat{g}_{\alpha\beta} dX^\alpha
dX^\beta + e^{-2\lambda} \left( f_1(z) dz^2 + f_2(z)
ds\Z{X_5}^2\right)\right].
\end{equation}
This metric ansatz solves all of the 10D vacuum Einstein
equations, when $f_1(z), f_2(z)\to {\rm const}$. The explicit
solution is given by
\begin{equation}\label{10d-soln2}
ds\Z{10}^2 = e^{-2{\cal K}|z|-z\Z{0}}  \left[
\hat{g}_{\alpha\beta} dX^\alpha dX^\beta + e^{-2\lambda}
e^{2z\Z{0}} \left( 2 {\cal K}^2 {dz}^2 + ds\Z{5}^2 \right)\right]
\end{equation}
with the scale factor
\begin{equation}
a(t)=\frac{1}{2} \exp\left(\mu t\,e^\lambda\right)+
\frac{\tilde{k} e^{-2\lambda}}{2\mu^2} \exp\left(- \mu
t\,e^{\lambda} \right),\quad \mu\equiv
\sqrt{\frac{4}{3}}\,e^{-z\Z{0}}.
\end{equation}
Here $z\ge 0$; there is some sort of ``brane" at the $z=0$
boundary of spacetime.

Let us make sure that there are no any irrelevant integration
constants that can be set to zero. To this end, one may introduce
a new symbol $H$ such that $\mu e^\lambda\to H$. Hence
\begin{equation}
a(t)=\frac{1}{2}\, e^{H t}+ \frac{\tilde{k}}{2 H^2}\, e^{- H t}.
\end{equation}
The metric that explicitly solves all of the 10d vacuum Einstein
equations is
\begin{equation}\label{10d-soln3}
ds\Z{10}^2 = e^{-2{\cal K}|z|- {\cal K}\Z{0}} \left(
\hat{g}_{\alpha\beta} dX^\alpha dX^\beta + \frac{4}{3H^2}\left( 2
{\cal K}^2 {dz}^2 + ds\Z{5}^2 \right)\right),
\end{equation}
where ${\cal K}\Z{0}$ is an integration constant. There are two
free parameters in the model, i.e. $H$ and ${\cal K}$, other than
the conformal factor $\exp(-{\cal K}\Z{0})$ multiplying the 10d
metric. In fact, one should not fix the parameter like ${\cal
K}\Z{0}$ arbitrarily; the choice ${\cal K}\Z{0}=0$ could
unnecessarily over-constrain the model (see below). Here one also
notes that there is no static limit of the above solution, i.e.
$H=0$ is not physical, irrespective of the choice $\tilde{k}=0$ or
$\tilde{k}\ne 0$.

A straightforward calculation yields
\begin{subequations}
\begin{align}
\sqrt{-g\Z{10}}=\frac{16\sqrt{2}|K|}{729 H^6}\, e^{-10K |z|
-5K\Z{0}}
\sin\theta\Z{1}\sin\theta\Z{2} \sqrt{\hat{g}\Z{4}},\\
R\Z{10}=e^{2K|z|+K\Z{0}} \left(\hat{R}_4 -12 H^2\right).
\end{align}
\end{subequations}
The 4d effective cosmological constant is given by $\Lambda_4= 6
H^2$. Note that in the above example the expansion parameter $H$
is not an independent or free parameter, it is rather related to
the overall size of the extra dimensions. From the above solution
we derive
\begin{equation}
M_{Pl}^2 \equiv \frac{M\Z{10}^8\times V_6^{\rm
w}}{(2\pi)^6}=\frac{M\Z{10}^8}{\pi^3}\times
\frac{8\sqrt{2}\,|{\cal K}|\,e^{-\,4{\cal K}\Z{0}}}{729\,H^6}
\int_{-z_1}^{z_1} dz\, e^{-8 {\cal K} |z|},
\end{equation}
Here $z\ge 0$; there is some sort of `brane' at the $z=0$ boundary
of spacetime. $M\Z{10}$ and $V_6^{\rm w}$ are, respectively, the
10d Planck mass and the warped volume of the internal 6d space.
The four-dimensional Planck mass is finite even in the limit
$z_1\to \infty$. The warping becomes stronger away from the brane.
In other words $e^{-8{\cal K} |z|}$ becomes exponentially small as
$|z|\to \infty$. This behavior is similar to that in braneworld
models. The larger is the number of extra dimensions, the stronger
would be the effect of warping in the transverse directions,
implying that the Kaluza-Klein particles could become more massive
as the number of dimensions increases.

For the solutions presented above, $\Lambda\Z{4}$ may be tuned to
be the present value of dark energy density by taking $H\sim
10^{-60} ~M_{\rm Pl}$. Then, with ${\cal K} \sim {\cal O}(1)$, we
get $M\Z{10} \gtrsim {\rm TeV} \sim 10^{-15}~M_{\rm Pl}$ for
${\cal K}\Z{0} \gtrsim 136$. The message is simple and clear: the
choice like ${\cal K}\Z{0}=0$ as {\it a priori} could easily lead
to some undesirable results.

While mathematically consistent it is physically more appealing to
generalise the present construction by considering the effects of
sources other than gravity or metric flux (i.e. the effect of a
non-zero Ricci scalar). One would view the above constructions as
local models, which could be easily embedded or capped off in
string theory leading to an honest de Sitter compactification by
introducing some known objects in string theory. The usual such
objects are background fluxes (or form fields), orientifold planes
and D7-branes with D3-brane charge induced by curvature on the
brane world-volume~\cite{GKP}.

The important ingredient in this present approach to understanding
of accelerating universes is the treatment of `brane as a
four-dimensional spacetime or physical universe and consideration
of non-factorizable background geometries, where the metric of the
four familiar dimensions is dependent of a coordinate in the extra
dimensions. Within this set-up and in spacetime dimensions $D\ge
5$, inflationary cosmology is possible with a wide class of
metrics and also with an arbitrary scalar curvature of the
internal space, if the number of extra dimensions $(D-4)\ge 2$. A
salient feature of the inflationary de Sitter solution given above
is that upon the dimensional reduction from $D$-dimensional
Einstein gravity one generates in the 4d effective theory a
cosmological constant-like term, which is nothing but
$\Lambda_4=6\mu^2 e^{2\lambda}$. The effective four-dimensional
theory is Einstein's general relativity supplemented with a
cosmological constant-like term.


\end{document}